\documentstyle[multicol,aps,psfig]{revtex}

\begin{document}

\newcommand{\la} {\langle}
\newcommand{\ra} {\rangle}
\newcommand{\ep} {\varepsilon}
\newcommand{\bu} {\bf u}
\newcommand{\de} {\delta}

\pagestyle{myheadings}
\markright{To be submitted to PRL}
\draft
\preprint{Submitted to Phys. Rev. Lett.}

\title{Clustering Kinetics of Granular Media in Three Dimensions}

\author{Shiyi Chen$^{1,2}$, Yuefan Deng$^{3}$, Xiaobo Nie$^{1}$ and 
Yuhai Tu$^{2}$}
\address{${}^{1}$Center for Nonlinear Studies,
Los Alamos National Laboratory, Los Alamos, NM 87545\\
${}^{2}$IBM Research Division, T. J. Watson Research Center,
P.O. Box 218, Yorktown Heights, NY 10598\\
${}^{3}$Department of Applied mathematics and Statistics, SUNY 
at Stony Brook, Stony Brook, NY 11794-3600}

\maketitle

\begin{abstract}

Three-dimensional molecular dynamics simulations of dissipative particles 
($\sim 10^6$) are carried out for studying the clustering
kinetics of granular media during cooling. The inter-connected
high particle density regions are identified, showing tube-like structures. 
The energy decay rates as functions of the particle density and the 
restitution coefficient are obtained. It is found that the probability 
density function of the particle density obeys an exponential distribution 
at late stages. Both the fluctuation of density and the mean cluster size 
of the particle density have power law relations against time during the 
inelastic coalescing process.

\end{abstract}
\pacs{83.10.Pp, 83.70.Fn, 81.05.Rm}

\begin{multicols}{2}
\narrowtext

To understand the fundamental differences of macroscopic behaviors between 
a regular gas and the inherently dissipative granular media is a
challenging problem. Recent studies \cite{jaeger} have 
revealed that
inelastic collision in granular media gives rise to a variety of complex
phenomena which are absent in classical ideal gas,  including the formation of 
clustering structures \cite{gold,brey,kud,deltour} and 
collapsing \cite{mcn,zhou}. 
To understand the clustering kinetics and to make the  
connection between the microscopic and macroscopic properties of granular 
media are not only crucial for constructing 
the constitutive equations of granular materials, but also important 
for various engineering 
applications, including packing \cite{eli}, size segregation \cite{william,fan}
and transport of granular materials\cite{jaeger,umbanhowar}. 

Cluster formation has been studied 
before. In particular, using a model of hard disks with dissipation,
Goldhirsch and Zanetti \cite{gold} found that the 
clustering particles form long string-like structure in two dimensions.
Experimentally, two dimensional cluster formation has 
been studied in a driven system by Kudrolli {\em et al.} 
\cite{kud} 
using a system consisting of small steel spheres rolling on a smooth 
bounded rectangular surface with moving side walls. They measured the 
probability density 
functions (PDF) of the particle density and the particle velocity for the 
clusters near the driven wall. However, most 
existing numerical simulations and experiments for clustering dynamics are 
carried out for two dimensions and only a small number of granular 
particles (in the order of 10,000) are used. For such a system,
it is difficult to obtain reliable statistics and  to
address problems connecting macroscopic description with microscopic dynamics. 

In this Letter, we present an analysis of three-dimensional clustering kinetics
of granular media in a free cooling condition using a recently developed 
three-dimensional
molecular dynamics (MD) code based on Massage-Passing-Interface (MPI). 
This parallel code makes it possible to simulate over one million 
dissipative particles in 
three dimensions
and therefore allows us to tackle problems which are difficult in 
previous studies, including the growth of fluctuation density and the
mean cluster size of the particle clusters. In this paper, we focus on 
studying the energy decaying dynamics, three-dimensional structures of clusters 
and the scaling kinetics of the particle density.  

The granular media in this paper is treated as a collection of identical 
spherical particles with radius $R$. The motion of 
each particle is simulated by Newton's second law while ignoring the 
angular motion. To specify the interaction forces in our MD simulation 
model, we assume that there are no interactions when particles are not 
in contact.  The following interaction forces are included when 
two particles, $i$ and $j$ overlap\cite{herrmann}, i.e., the distance
$|{\bf r}_{ij}|$ (${\bf r}_{ij} \equiv {\bf r}_j - {\bf r}_i$) is smaller 
than $2R$: (1) An elastic restoration force: 
$f_{el}^{(i)} = Y m_i(|{\bf r}_{ij}| - 2R){\bf r}_{ij}/|{\bf r}_{ij}|$;
(2) dissipations due to the inelasticity of the collision:
$f_{diss}^{(i)} = - \gamma m_i{\bf v}_{ij}^{n}$
and $f_{shear}^{(i)} = -\gamma_s m_i{\bf v}_{ij}^{t}$. 
Here $Y$ is the Young's modulus, $m_i \sim R^3$ is the mass of particle
$i$; ${\bf v}_{ij}^n = ({\bf v}_{ij}\cdot{\bf r}_{ij}){\bf r}_{ij}/|{\bf r}_{ij}|^2$ 
and ${\bf v}_{ij}^{t} = {\bf v}_{ij} - {\bf v}_{ij}^n $ 
are the projections of the relative velocities ${\bf v}_{ij}$ 
($\equiv {\bf v}_i - {\bf v}_j$) on the ${\bf r}_{ij}$ direction and 
the tangential direction, respectively; 
$\gamma$ and $\gamma_s$ are the dissipation 
coefficients for the relative motion of particles in the
normal and tangential directions. 

Initially particles are distributed uniformly in a cubic box with a small 
random perturbation. The initial particle velocities are drawn from an
isotropic Gaussian distribution with 
variance $\sigma$.
For simplicity, in this paper we only report results for 
$\sigma =4$  and the Young's modulus $Y=10^5$.  The radius of granular
particles, $R$,  and the mass of the particles, $m$, are normalized to 
be $1$. A second-order scheme is used for the time integration 
of the motion of particles and simulations are carried out in a periodic box 
of size $L^3$.  Because of the large value of Young's modulus $Y$, our time 
step $\Delta t$ has to be set to  
$\Delta t=4.0\times 10^{-4}$, small enough to resolve the full 
collision process.  The particle volume 
fraction $\alpha$ is defined as $\alpha  = 4\pi N/3L^3$, where $N$ is
the particle number in the system.  The restitution coefficient, $r$, 
defined as the ratio of the relative speeds after collision and 
before collision, can be determined as: $r = \sqrt{(r_n^2  + 2r_s^2)/3}$, 
where $r_n = \exp(-\pi/\sqrt{\omega^2/\eta^2 -1})$, 
$r_s = r_n^{2\gamma_s/\gamma}$,
$\omega = \sqrt{Y/m}$ and $\eta = \gamma/2m$ \cite{note1}. 

\bigskip
{\psfig{file=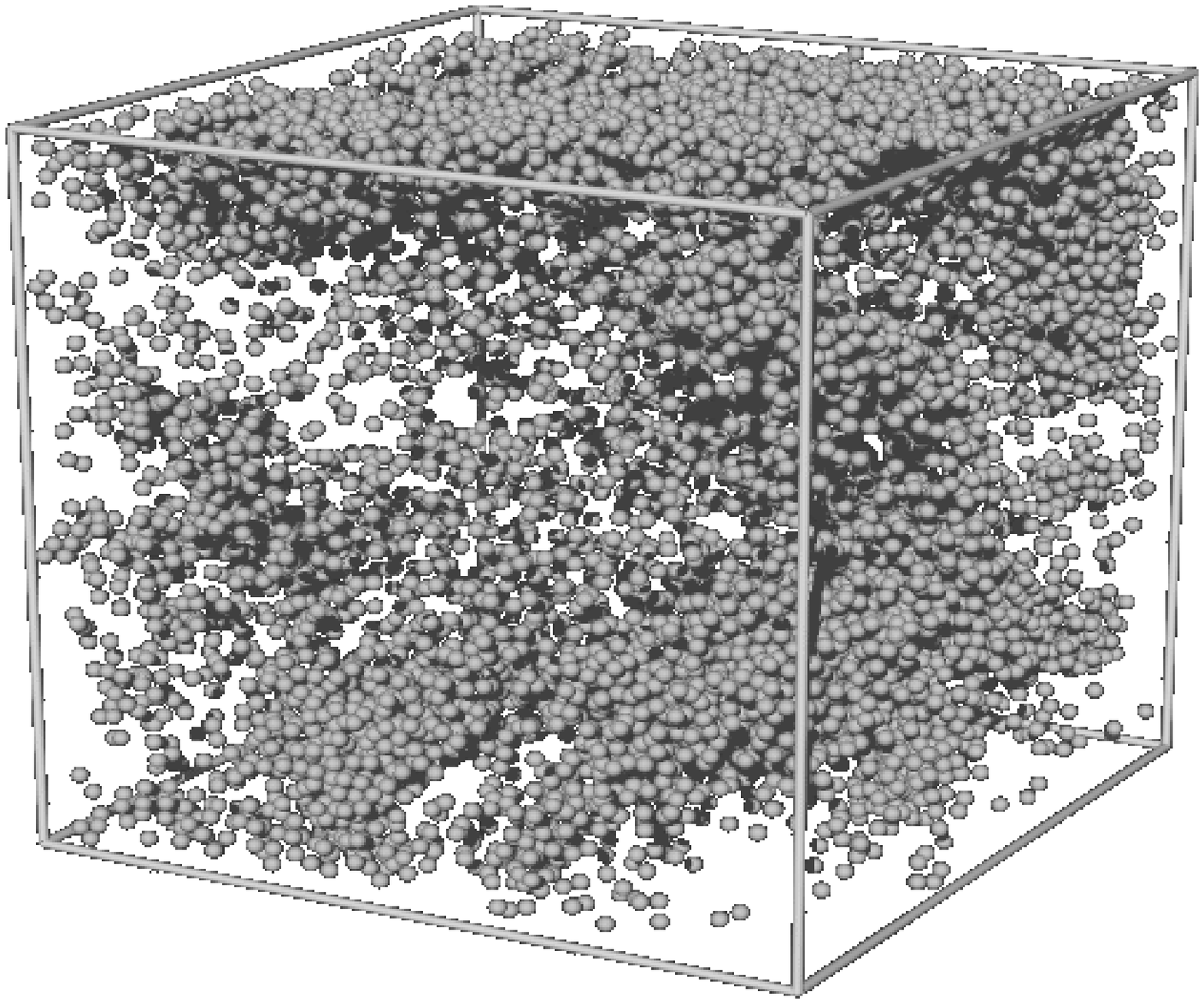,width=220pt}}
\noindent
{\small FIG.~1. A typical particle configuration at late stages of clustering 
formation for parameters $\alpha=0.074$ and $r=0.6$.
The simulation is for $10^6$ particles, but only
one-sixty-fourth of the simulation domain is shown.}

\bigskip
\psfig{file=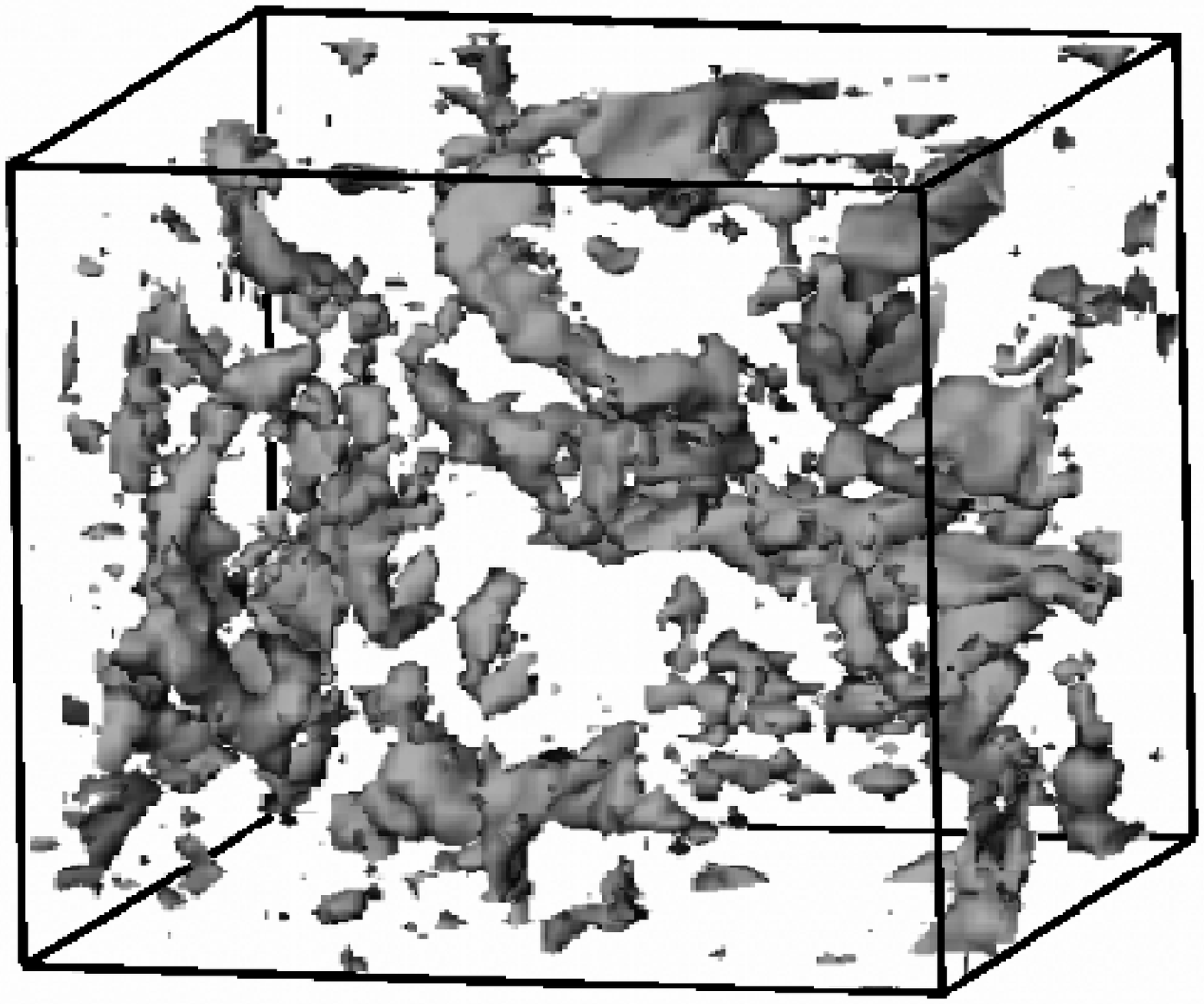,width=220pt}
\noindent
{\small FIG.~2. Iso-surface of the particle density for $\rho = 3.5 \rho_m$.
The resolution in this plot is $50^3$ and the simulation was carried out
for $L = 384$.}
\bigskip

In Fig.~1, we show a snapshot of a three dimensional particle 
distribution in a cooling granular system 
at late stage $t = 1000$ from a simulation of one million particles
(the initial characteristic collision time is of $O(1)$).  
The parameters in this calculation are 
$\alpha=0.074$, $\gamma =102$ and $\gamma_s = 51$, which result in $r = 0.6$. 
For clarity, only one-sixty-fourth of the whole simulation 
domain is shown in this plot. To calculate the local particle 
density $\rho({\bf x})$, we equally divide the whole space
into small sub-boxes and average the particle number
in each box.  In Fig.~2, we
present iso-surfaces of the particle density function $\rho({\bf x})$ for
$\rho = 3.5 \rho_m$, where $\rho_m$ is the mean density of the whole space.
From Fig.~1 and Fig.~2,  it is evident that the particle clusters
are formed and are more or less spatially inter-connected. In addition,
the cluster structures are tube-like, which is to be compared with
the string-like particle structure in two dimensions\cite{gold}. 
We have also calculated fractal dimensions for different regions using the 
box-counting method\cite{fra}.  We found that the fractal dimension 
$d_{f} \simeq  1.3$ for $\rho \ge 3.5 \rho_m$, in a qualitative agreement
with the observed tube-like structure in Fig.~2,  and 
$d_{f}\simeq 2.2$ for $\rho \ge \rho_m$. 

In a cooling system, the forms of the particle motion fall roughly into two
categories: free streaming and collision. During the streaming process, 
the speed of each particle is constant. The collision step decreases the 
speed of particles involved in the collision due to dissipation and reduces
the energy. Based on a kinematic argument, it has been shown 
in \cite{gold,McNamara93} that if the spatial particle distribution is uniform
for which the mean free path is proportional to $E^{1/2}$,
and the local velocity distribution is random which leads to strong collision
with loss of energy proportional to $E$, then the total energy of the system,
$E(t) = 1/2 \sum_i {\bf v}_i^2$, decays as: 
\begin{equation}
E(t) = \frac{E(0)} { (1 + t/t_c)^\beta}, 
\end{equation}
where $t_c$ is a crossover time, and the exponent $\beta$ equals $2$.

Our simulations show that in general the above equation can be used to 
describe the decay of the total energy during the particle 
clustering process, but $\beta$ depends on the properties of the granular 
media, in particular, the particle volume fraction and the restitution 
coefficient. Fig.~3a shows the energy decay exponent $\beta$ as a function of
the particle volume rate $\alpha$ for the restitution coefficient
$r = 0.43$.  In the inset, the total energy versus time for
a typical case is shown. One can see that a power law decay can 
be identified for the late stage.  Similar results are 
shown in Fig.~3b for $\beta$ against the restitution coefficient $r$
with $\alpha = 0.09$. The results shown here give a quantitative 
description of the dependence of the decay exponent on various material 
parameters: in the case of low density or large restitution, 
the particle distribution in the system is almost uniform and 
the resulting decay exponents equal $2$, in a good agreement with 
the heuristic argument in \cite{gold,McNamara93}. 
The smaller decay exponents are seen for systems 
with high densities and small restitutions, where clustering occurs. 
In this circumstance, the speed of particles in a cluster is 
approximately the same and the collision frequency is relatively low, 
leading to less energy dissipation and smaller energy decay rate. 

The decrease of the energy decay rate has been studied previously
in \cite{gold,deltour} using a linear hydrodynamic analysis
based on phenomenological macroscopic equations. 
It is argued that for a system with particle clusters, the hydrodynamic
modes have low decay rates. Quantitatively, it is interesting to notice 
that the energy decay exponent $\beta$ 
is round $1.4\sim 1.5$ in the high density and small restitution 
regimes. To understand this nontrivial exponent,
we decompose the total energy: $E(t) = E_k(t) + E_f(t)$, 
with $E_k(t)$ and $E_f(t)$ being
the local spatially averaged kinetic energy and the fluctuation energy,
respectively. We argue that in the high density or small restitution case, 
due to the formation of particle clusters, the motion of the 
clustered particles dominates, therefore the decay rate of $E(t)$ follows 
directly the macroscopic hydrodynamic decay of $E_k(t)$:
\begin{eqnarray}
E_{k}(t)&\sim& \int |\vec{v}(\vec{k},t=0)|^2\exp(-2\nu |\vec{k}|^2 t)
d^{D}\vec{k}\nonumber\\
&\sim& t^{-D/2}, 
\end{eqnarray}
where $\nu$ is some effective viscosity and $D$ is the spatial dimension.
In deriving (2), We have assumed that the initial velocity fluctuation 
$\vec{v}(\vec{k},t=0)$ is independent of $\vec{k}$ for small 
$|\vec{k}|$ (certainly true for the initial velocity distribution).
For $D=3$, Our numerical results of $\beta \sim 1.5$ in the clustering regime 
agree well with Eq. (2). To further verify our prediction, 
we have also carried out 2D simulations (not shown here). We found that the 
energy decay rate at high densities is indeed close to $1$, while the
exponent $\beta$ for very low density remains $2$. 

\bigskip
{\psfig{file=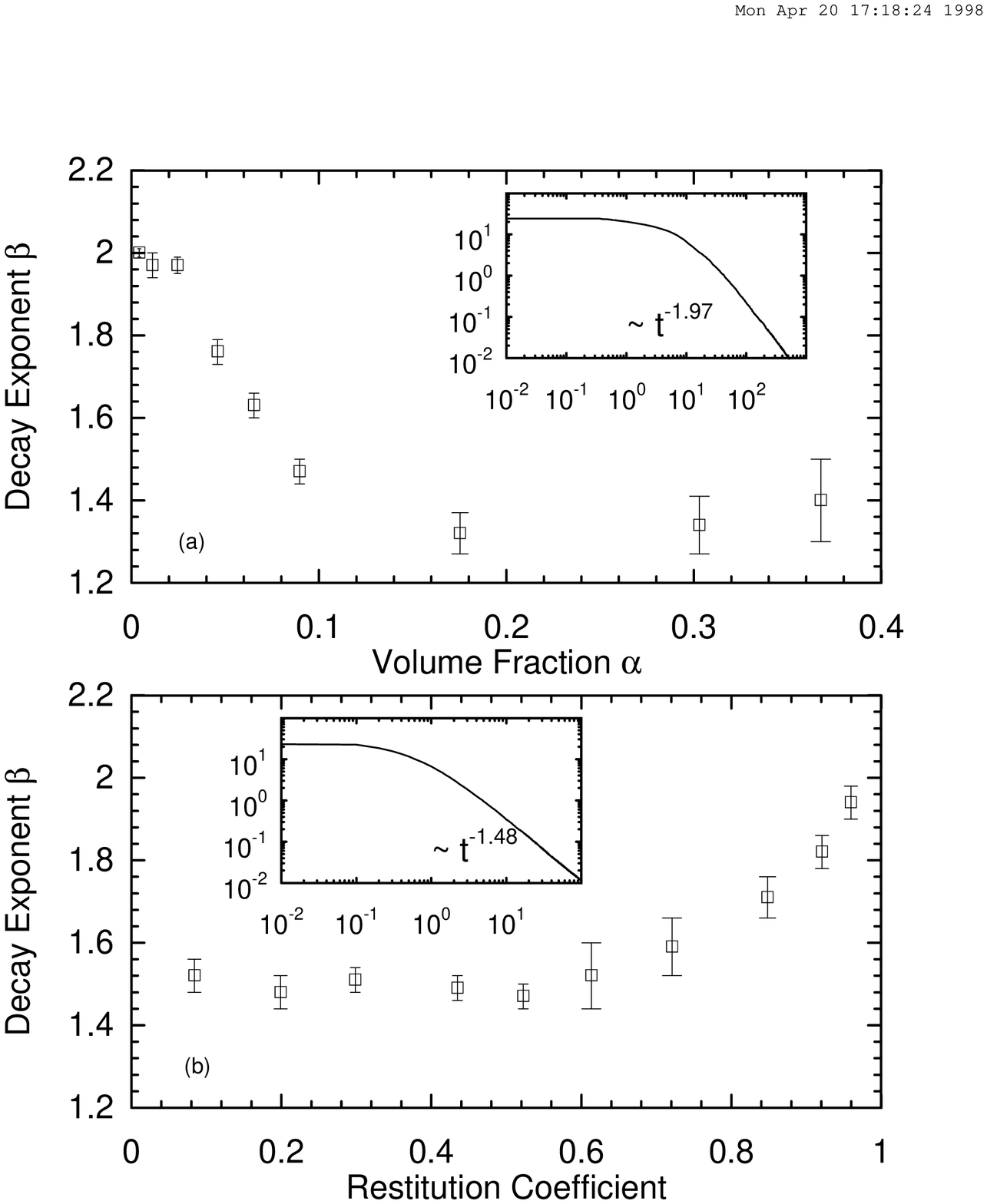,width=220pt}}
{\small FIG.~3. The energy decay exponent $\beta$
as functions of the particle volume fraction (a) and
the restitution coefficient (b).
The inserted pictures show the typical energy decays against time
for (a) $\alpha = 0.011$; (b) $r =  0.20$.}
\bigskip

In the absence of clustering, the spatial particle distribution
should be uniform and the PDF of the particle density 
is given by a $\delta$-function. On the other hand, if clustering 
occurs, particles coalesce to certain regions while leaving others 
void, leading to the disparity of particle density distribution. 
The study of the time dependence of the PDF of the particle density 
greatly enhances our knowledges of the clustering process and the 
characteristics of the clusters. It is interesting to note that the 
shape of the PDF is correlated
with the energy decay exponent $\beta$. For $\beta \simeq 2$, the
corresponding PDF of the density shows very little spreading, signaling
the non-existence of clustering in that parameter range, whereas in the
clustering parameter regime (e.g., the parameters used in Fig.~4),
$\beta$ is close to $1.5$. In Fig.~4, we present the PDF of the density 
function $\rho$ at different times with the same parameters as in Fig.~1.
The statistics are obtained using $32^3$ sub-boxes.
The density is normalized by the average density $\rho_m$. 
The PDF at $t=0$ is centered at one narrow density interval due to
our specific initialization of the particle distribution.
As the systems evolves, the PDF of the density function
widens, implying the non-uniformity of the spatial
particle distribution and the formation of clusters. Finally, the system 
becomes totally spread with more samples at lower density and
fewer at high density. The PDFs at late stages can be approximated by an 
exponential function, indicating the existence of all possible
densities. No local peak of PDF at high density was observed, at least for the 
simulation times we have reached, implying no specific characteristic
density scale.  Finally, in the inset of Fig.~4, we show 
the density fluctuation $\delta \rho(t)$ as a function of time, which 
follows a power law for almost three decades in time:
\begin{equation}
\delta \rho(t)^2\equiv \langle (\rho - \rho_m)^2 \rangle \sim t^{0.55}\;\; .
\end{equation}

\bigskip
{\psfig{file=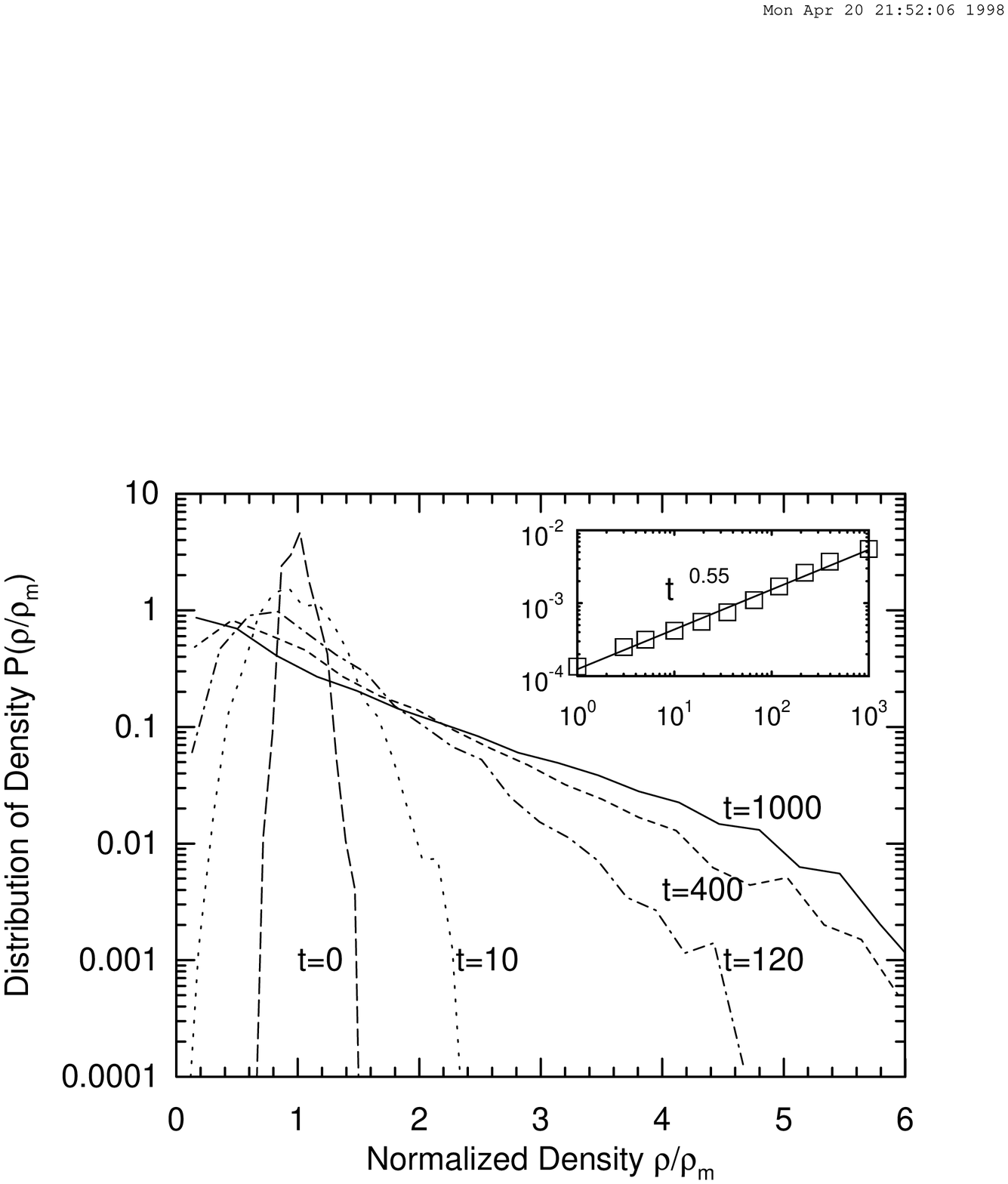,width=220pt}}
{\small FIG.~4. Normalized PDFs of the particle density at 
different times. The inserted picture shows that the fluctuation of the
density function grows as a power function of time.}
\bigskip

To study the particle density correlation, we analyze
the power spectra of the particle density function: 
$E_{\rho}(K) = \sum_{K-1/2 < k < K+1/2} \rho(k)\rho^{*}(k)$, where
$\rho(k)$ is the particle density in Fourier space. 
In the inset of Fig.~5, $E_{\rho}(K)$ as a function of $K$ at 
different times are shown. Two features in the power spectra can 
readily be seen.  First, as the time evolves, the peak value of the 
power spectra increases, in agreement with the inset in Fig.~4 due to the fact
 $\delta \rho^2 = \int_K E_{\rho}(K) dK$. Second, if we denote $K_{max}(t)$
for the wavenumber at which $E_{\rho}(K)$ has the maximum value,
we can see that $K_{max}(t)$ decreases as a function of time, indicating 
the increase of the typical cluster size.
To quantify this behavior, we define a mean cluster size, $\lambda(t)$ by: 
$\lambda(t) = \int_{K} E_{\rho}(K) dK/ \int_{K} E_{\rho}(K) K dK$. 
In Fig.~5, we present $\lambda(t)$ as a function of time for the 
times when particles form clusters.  Again, a power law dependence
of $\lambda(t)$ on $t$ is observed over two decades in time\cite{foot2}:
\begin{equation}
\lambda(t) \sim t^{\zeta}\;\;, 
\end{equation}
where $\zeta$ is the growth exponent
for the mean cluster size and is close to $0.13$ for the
given case.  From other simulations not shown here, we note that $\zeta$ has a 
week dependent on the particle volume fraction and increases slowly with
increasing of $\alpha$. 

\bigskip
{\psfig{file=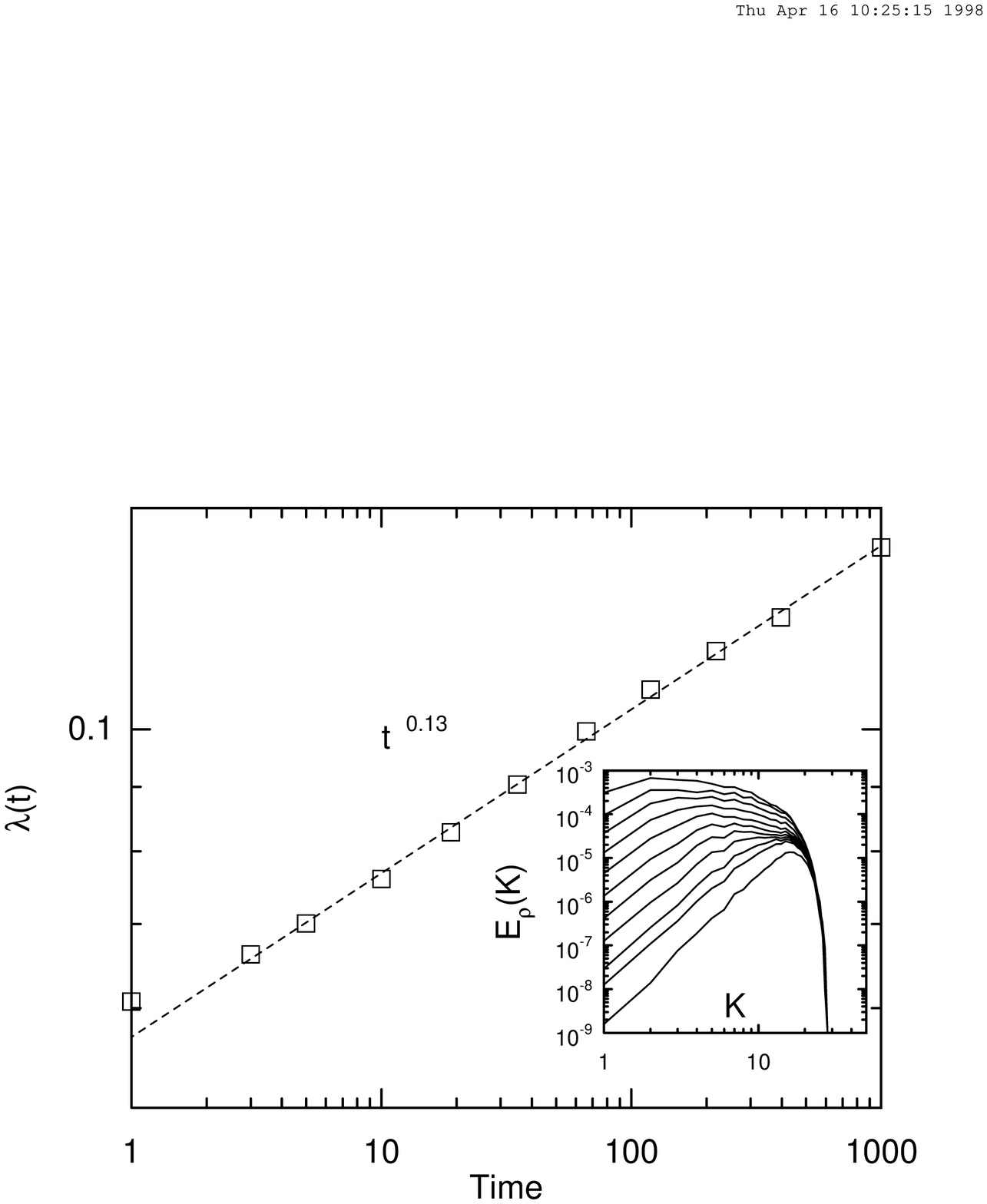,width=220pt}}
{\small FIG.~5. The mean size of particle cluster as 
a function of time. In the inset shows the power spectra  
$E_{\rho}(K)$ at different times. }
\bigskip

To summarize, we have studied the clustering phenomenon
in the cooling process of three dimensional granular media. 
Large spatial density fluctuation is observed due to inelastic
collision and the formation of the clusters strongly affects the 
energy decay rate.  It is found that the total energy 
has a power law decay against time at late stages: 
 $E(t) \sim t^{-\beta}$ with $\beta \sim 1.5$ in the parameter 
regimes where clustering forms. 
The value of $\beta$ crosses over to its mean field value 
$\beta_{MFT}=2.0$ for systems with large restitution constant $r$ or
small density $\alpha$ where the particle clustering is absent. 
By making simple analogy to the decay of kinetic energy in fluids, we 
relate the value of $\beta$ in the clustering parameter regime to the spatial 
dimension $D$: $\beta=D/2$. We observe that clusters in three dimensions
form tube-like structures for high particle density regions at late stages. 
A preliminary study has revealed that the structure might be
fractal. We have also 
studied the dynamical properties of clusters. It is found that 
the density fluctuation and the cluster size both 
obey power law scalings against time with non-trivial scaling 
exponents. We have no theory to explain the  observed scaling dynamics
and it is our hope that the present results would stimulate further 
studies in understanding these intriguing behaviors. 

We thank Eli Ben-Naim and Gary D. Doolen for useful discussions. 
Numerical simulations were carried out at the Center for Scalable 
Computing Solution at IBM T. J. Watson Research Center using the SP machines.

\end{multicols}


\begin{references}

\bibitem{jaeger} H. M. Jaeger, S. R. Nagel and R. P. Behringer, Physics
Today, 32-38, April, 1996; H. M. Jaeger, S. R. Nagel and R. P. Behringer,
Rev. Mod. Phys. {\bf 68}, 1259 (1996). 

\bibitem{gold} I. Goldhirsch and G. Zanetti, Phys. Rev. Lett. {\bf 70}, 
1619 (1993). 

\bibitem{brey} J. J. Brey, M. J. Ruiz-Montero and D. Cubero, Phys. Rev. E.
{\bf 54}, 3664 (1996).

\bibitem{kud} A. Kudrolli, M. Wolpert and J. P. Gollub, Phys. Rev. Lett.
{\bf 78}, 1383 (1997).

\bibitem{deltour} P. Deltour and J.-L. Barrat, {\em Quantitative study
of a free cooling granular medium}, preprint, (1997).

\bibitem{mcn}S. McNamara and W. R. Young, Phys. Rev. E, {\bf 50}, R28 (1994). 

\bibitem{zhou} T. Zhou and L. P. Kadanoff, Phys. Rev. E, {\bf 54}, 623 (1996). 

\bibitem{eli}E. R. Nowak, J. B. Knight, E. Ben-Naim, H. M. Jaeger, and S. R. 
Nagel, Phys. Rev. E {\bf 57}, 1971 (1998); E. Ben-Naim, J. B. Knight, E. R. Nowak, 
H. M. Jaeger and S. Nagel, Phys. Rev. E. preprint (1998). 

\bibitem{william} J. C. Williams, Power Technol. {\bf 15}, 245 (1976). 

\bibitem{fan}L. T. Y. M. Fan and F. S. Lai, Powder Technol. {\bf 61}, 255 (1976).

\bibitem{umbanhowar} P. B. Umbanhowar, F. Melo and H. L. Swinney, 
Nature. {\bf 382}, 793(1996).

\bibitem{herrmann} H. J. Herrmann, {\em Computer simulation of granular media}, in
{\em Disorder and Granular Media}, ed. D. Bideau and A. Hansen, 1993 Elsevier 
Science Publishers.

\bibitem{note1} The connection between the restitution coefficient $r$ and
the parameters of the granular media can be obtained through the following 
calculation: Let $x(t)$ and $y(t)$ be the relative coordinates
along the ${\bf r}_{ij}$ direction and the tangential direction during 
two particle collision, we have: $m \ddot{x} + \gamma \dot{x} + Y x=0$ and
$m \ddot{y} +\gamma_s y=0$. The solution of these equations 
gives the relative velocities after collision in these two directions: 
$-r_n {\bf v}_{ij} cos(\theta)$ and $r_s {\bf v}_{ij} sin(\theta)$. 
Then $r^2 = r_n^2 cos^{2}(\theta) + r_s^2 sin^{2}(\theta)$.
The average of $r^2$ over the relative angle $\theta \in (0, \pi/2)$
leads to $r^2 = (r_n^2 + 2r_s^2)/3 $.

\bibitem{fra}Jens Feder, {\em Fractal}, 1988 Plenum Press, New York.

\bibitem{McNamara93}S. McNamara, Phys. Fluids A. {\bf 5}, 12 (1993).

\bibitem{foot2} The deviation from the power law behavior 
at the early stage $t < 5$ could be due to initial particle-particle
interaction, which is different from the cluster-cluster interaction
of the late stage growth. 
 

\end{references}
\end{document}